\documentclass[english]{article}
\usepackage[utf8]{inputenc}
\usepackage[T1]{fontenc}
\usepackage{babel}
\usepackage{amsmath}
\usepackage{graphicx}
\usepackage{fancyhdr}
\usepackage[hidelinks]{hyperref}
\usepackage{soul,color}
\usepackage{caption}
\usepackage{float} 
\usepackage{geometry}
\usepackage{booktabs}
\usepackage{multirow}
\begin{document}
	
	\title{Multiscale dynamic human mobility flow dataset in the {U.S.} during the COVID-19 epidemic}
	
	\author{Yuhao Kang\textsuperscript{1}, Song Gao\textsuperscript{1{*}}, Yunlei Liang\textsuperscript{1}, Mingxiao Li\textsuperscript{1,2,3}, Jinmeng Rao\textsuperscript{1}, Jake Kruse\textsuperscript{1}}
	
	\date{August, 2020}
	
	\maketitle
	\thispagestyle{fancy}
	1. GeoDS Lab, Department of Geography, University of Wisconsin-Madison, WI 53706, United States; 2. Institute of Geographic Sciences and Natural Resources Research, Chinese Academy of Sciences, Beijing 100101, China; 3. School of Architecture and Urban Planning, Shenzhen University, Shenzhen 518061, China\\
	{*}Corresponding Author: Song Gao (song.gao@wisc.edu)

	\begin{abstract} 
		
		\footnote{A preprint version. Final corrected proof is available on Scientific Data: \url{https://www.nature.com/articles/s41597-020-00734-5}}Understanding dynamic human mobility changes and spatial interaction patterns at different geographic scales is crucial for assessing the impacts of non-pharmaceutical interventions (such as stay-at-home orders) during the COVID-19 pandemic. 
		In this data descriptor, we introduce a regularly-updated multiscale dynamic human mobility flow dataset across the United States, with data starting from March 1st, 2020.
		By analysing millions of anonymous mobile phone users’ visits to various places provided by SafeGraph, the daily and weekly dynamic origin-to-destination (O-D) population flows are computed, aggregated, and inferred at three geographic scales: census tract, county, and state.
		There is high correlation between our mobility flow dataset and openly available data sources, which shows the reliability of the produced data. 
		Such a high spatiotemporal resolution human mobility flow dataset at different geographic scales over time may help monitor epidemic spreading dynamics, inform public health policy, and deepen our understanding of human behaviour changes under the unprecedented public health crisis. 
		This up-to-date O-D flow open data can support many other social sensing and transportation applications.
	\end{abstract}
	
	\section*{Background \& Summary}
	The outbreak of the novel coronavirus disease SARS-CoV-2 (also known as COVID-19) in December 2019 has become a global threat to public health and human societies.
	Thus far, more than 25 million people have been infected by the virus with more than eight hundred thousand death cases globally \cite{dong2020interactive}.
	To contain the transmission of the COVID-19, social distancing has been proved as the most effective non-pharmaceutical intervention \cite{Haushofereabb6144,lai2020effect,chen2020mitigating}, and governments have applied various policies to reduce human mobility and restrict large gatherings, such as regional lockdowns \cite{askitas2020lockdown}, stay-at-home orders \cite{wang2020covid, gao2020mobile}, and travel restrictions \cite{liu2020synchronized, chinazzi2020effect}.
	Tracking dynamic human mobility changes and spatial interaction patterns is therefore a prerequisite for measuring the effects of human mobility and interventions on predicting the virus spread \cite{ghader2020observed, buckee2020aggregated}. Several recent works have employed human movement flow matrices in understanding spatial interaction changes and social impact, and enabling network-based epidemic models to project the numbers of COVID-19 infected population in different countries such as China, Japan, Italy, France, Chile, and UK \cite{lai2020assessing,jia2020population,tian2020impact,li2020substantial,gatto2020spread,measuring2020chile, cintia2020relationship,bonato2020mobile,pullano2020population, yabe2020non,galeazzi2020human}, which requires up-to-date inbound and outbound human movement flow information. 
	However, there is no such openly and timely updated human movement origin-to-destination (O-D) flow matrix data at a fine spatiotemporal resolution available in many other countries where researchers can only use historical O-D survey data and other proxies as a compromise \cite{pei2020differential}.
	
	Human mobility has been widely studied in multiple disciplines such as geography, transportation, urban planning, physics, computer sciences, and public health \cite{barbosa2018human}. It reflects patterns about how people move from place to place and serves as an indicator of human behaviour and underlying socioeconomic environments.
	With the rapid development of information and communication technologies (ICT) and GPS embedded devices, large-scale mobile phone data provides an unprecedented opportunity in tracking human trajectories, which benefits research about human mobility patterns.
	Existing studies have used such data to investigate basic laws governing human movements \cite{gonzalez2008understanding, dong2020understanding}, model regional transportation connectedness and economy \cite{li2020estimation,boarnet2017urban}, describe daily commuting flows \cite{hincks2010spatial}, compute urban vibrancy \cite{jia2019measuring,tu2017coupling}, inform public health policy \cite{lai2020effect,li2020integration,jia2020population}, and understand spatial interaction patterns \cite{gao2013discovering,liu2014uncovering,ratti2006mobile}.
	
	As pointed out by several recent studies \cite{gao2020mapping,franch2020spatial,buckee2020aggregated, oliver2020mobile}, human mobility data plays a key role and serves as data foundation in the fight against the COVID-19 pandemic.
	Although companies such as Descartes Labs (\url{https://www.descarteslabs.com/mobility/}) \cite{descarteslabs2020mobility}, Apple (\url{https://covid19.apple.com/mobility}), Google (\url{https://www.google.com/covid19/mobility/}) \cite{aktay2020google}, and Facebook (\url{https://dataforgood.fb.com/docs/covid19/}) \cite{kuchler2020geographic}, have released a set of near real-time mobility-related open datasets for monitoring human mobility changes and social distancing behaviour during the COVID-19 period, these datasets are lacking in three respects.
	First, human mobility flow matrices, which describe movement patterns from origin geographic units to destination regions, are often unavailable, even though such O-D paired flow matrices are incredibly valuable for epidemic transmission modeling and spatial-social interaction measurements \cite{li2020substantial,chen2020mitigating, kraemer2020effect, buckee2020aggregated,holtz2020interdependence}. For instance, only aggregated mobility indices (such as median travel distance and foot-traffic) at a specific region is provided by the Descartes Labs, Apple, and Google mobility datasets.
	Second, there is a lack of fine-resolution datasets regarding the privacy-accuracy trade off. 
	Most openly available datasets aggregate mobility patterns to the state, county or city scale, while higher spatial resolution (e.g., census tract) datasets, which provide more detailed human mobility patterns, are necessary to more accurately characterize heterogeneous human mobility changes within cities or intra-counties during the pandemic \cite{oliver2020mobile}.
	Third, the mobile phone or other sensor data-driven mobility patterns often only provide a sample (e.g., 10\%) of the entire population. There is no entire-population-level estimated flow data.
	
	To address the limitations of existing mobility databases, we introduce an openly available dataset that provides an estimation of dynamic population flows at multiple spatial scales (at the census tract, county, and state) and temporal resolutions (daily and weekly) across the U.S. during the COVID-19 pandemic considering the findability, accessibility, interoperability, and reusability of data \cite{wilkinson2016fair}.
	A similar mobility dataset has been released by researchers from Italy \cite{pepe2020covid}, while we focus on the mobility in the U.S.. 
	The O-D format dataset is generated by tracking millions of anonymous mobile phone users' trajectories collected by SafeGraph \cite{safegraph}.
	When producing the dataset, great efforts were taken to protect personal privacy by aggregating to various geographic scales so that individual information cannot be traced.
	Other public datasets, such as the the American Community Survey (ACS) commuting flows and the Descartes Lab COVID-19 mobility dataset are then compared to illustrate the reliability of the produced dataset.
	Such an up-to-date dataset can be a useful supplement in human mobility observation. It can be used not only in the fight against the COVID-19 pandemic, but also to benefit other researches and applications such as emergency response \cite{song2014prediction, huang2015geographic}, urban planning \cite{bassolas2019hierarchical}, and population migration \cite{hatton2005global}.

	\section*{Methods}
\begin{figure}[H]
		\centering
		\includegraphics[width=\textwidth]{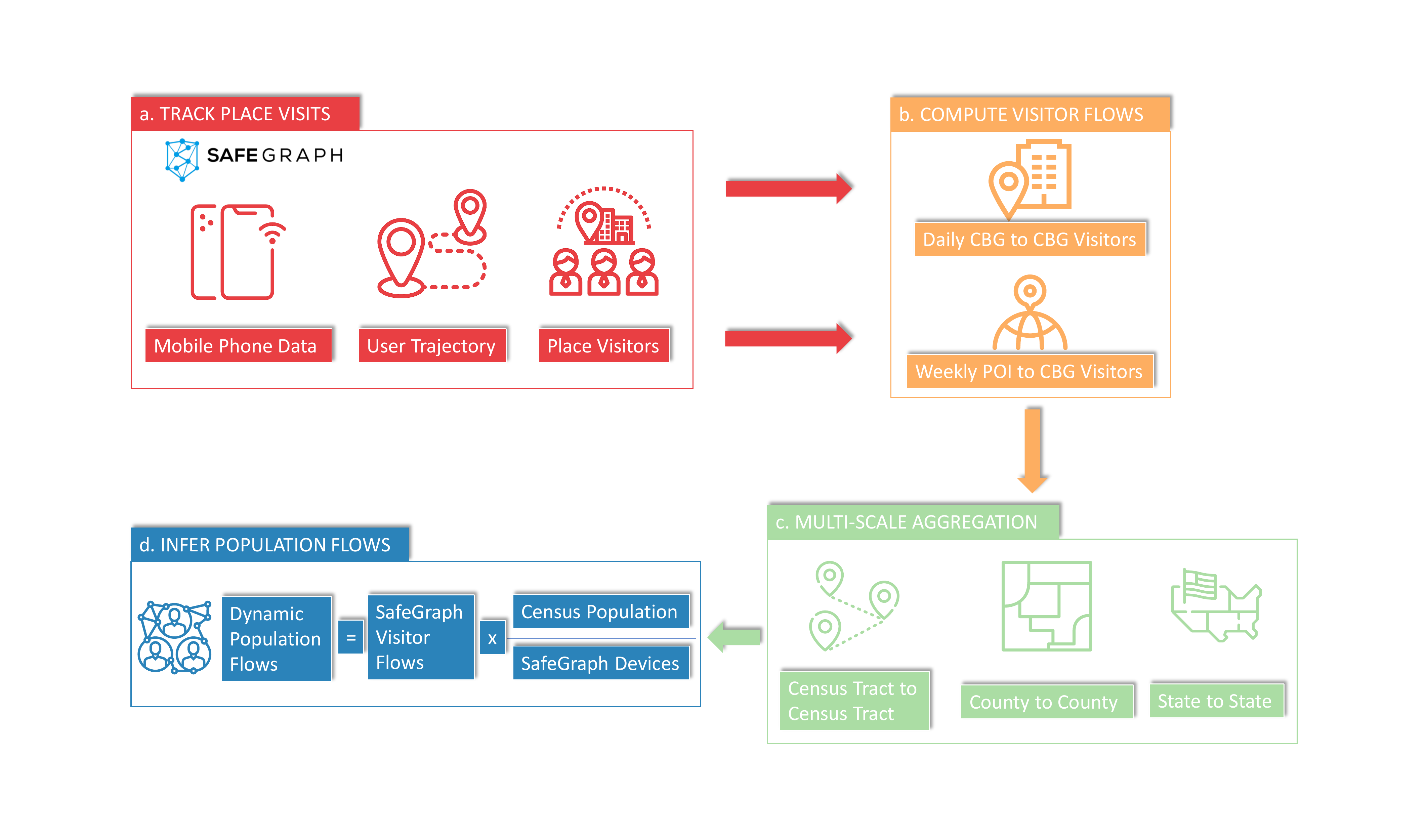}
		\caption{The data processing framework for the mobility flow dataset production: (a) track place visits; (b) compute visitor flows; (c) multi-scale aggregation; and (d) infer population flows.}
		\label{fig:framework} 
	\end{figure}

	Figure \ref{fig:framework} illustrates detailed processing steps for how this human mobility flow dataset is generated.
	The dynamic population O-D flow matrices are estimated using mobile phone location data provided by SafeGraph and demographic data retrieved from the ACS.
	Based on millions of anonymous mobile phone user visits to various places tracked by SafeGraph, two types of visitor flows, namely daily census block group (CBG) to CBG visitors and weekly CBG to point of interest (POI) visitors are computed, respectively.
	After spatially joining the place visitors to the administration regions, the visitor O-D flows are computed at three different spatial scales: census tract, county, and state, which are used to provide multi-scale views of human mobility and spatial interactions patterns between different places.
	Since the number of mobile phone users detected by SafeGraph is about 10\% sample of the entire population \cite{safegraphbias}, we further employ the ACS population data with mobile phone data samples to infer the population level of dynamic O-D flows.
	
	\subsection*{Track Place Visits}
	The place visitor patterns are retrieved from the SafeGraph COVID-19 Data Consortium \cite{safegraph}.
	Millions of anonymous GPS pings collected from numerous mobile applications are tracked and then cleaned to remove noise. 
	Then, users' home places are estimated and aggregated (e.g., at the level of a CBG), and those users' visits from home places to POIs are tracked.
	POIs are the primary venue for tracking place foot-traffic by SafeGraph, while CBG is one of the fine-resolution geographical units the United States Census Bureau used for publishing demographic and socioeconomic data.
	A home place of a user refers to his/her most common nighttime location during the last six weeks. 
	For each day, GPS pings of each device are clustered and only those clusters during nighttime hours (6pm - 7am local time) are kept.
	The CBG with the most clusters in that day is recorded.
	Based on this, the most frequent CBG over the last six weeks that reflects the primary nighttime location is used as the ``home location'' for each user.
	By aggregating home places to CBGs, user privacy can be protected as no individual records can be traced and accessed \cite{safegraphhome}.
	
	Active users' visits to POIs are produced with a similar strategy.
	Using several clustering methods such as density-based spatial clustering for applications with noise (DBSCAN)\cite{ester1996density}, GPS pings are grouped together in which each cluster contains a set of potential POIs and associates with CBGs.
	The best place for a given cluster is classified by performing machine learning methods involving several entangled features.
	Thereby each user's visits from home place to various POIs and CBGs are identified \cite{safegraphwhite}.
	
\begin{figure}[H] 
		\centering
		\includegraphics[width=\textwidth]{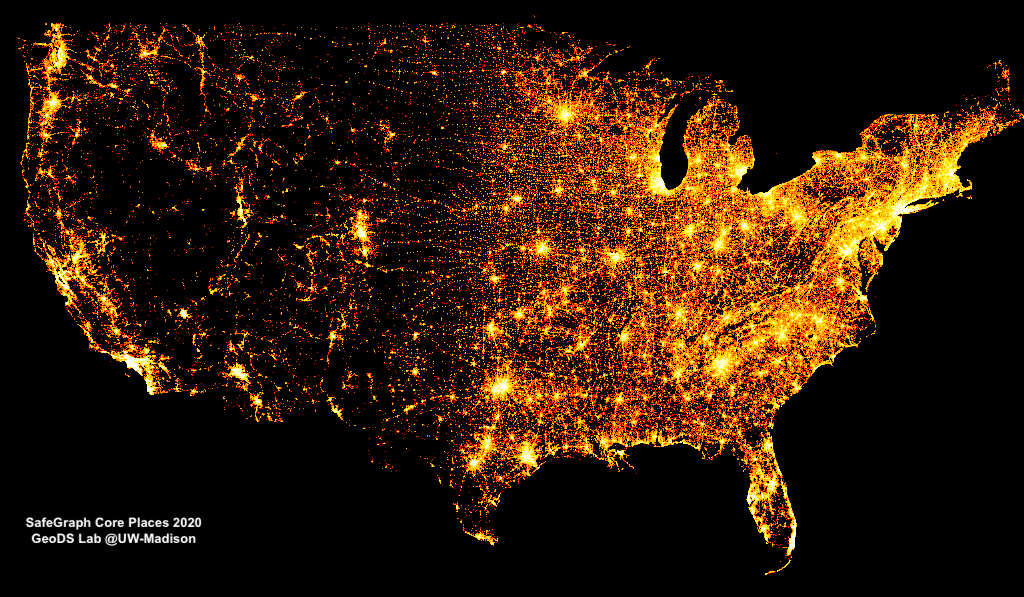}
		\caption{Spatial density distribution of places collected by SafeGraph across the whole United States; the visualization is created using the DataShader package, Python 3.7.}
		\label{fig:place_distribution}
	\end{figure}

	In total, there are more than 5 million POIs stored in the database, as well as more than 220 thousand CBGs retrieved from the ACS.
	The spatial density distribution of POIs across the Contiguous United States is mapped in Figure \ref{fig:place_distribution}, which shows that places cluster in major cities and are generally located along streets.
	The more places, the brighter the region in the map.
	
	\subsection*{Compute Visitor Flows}
	Two major human mobility flow metrics are employed in data production, and are denoted as \textit{daily CBG to CBG visitor flows} and \textit{weekly CBG to POI visitor flows}.
	In the \textit{daily CBG to CBG visitor flows} metric, each row contains an origin CBG and a destination CBG, as well as the number of mobile phone-based visitor flows from the origin CBG to the destination CBG.
	Every day, the number of unique mobile phone users who live in the origin CBG and visits to the destination CBG are recorded.
	More specifically, GPS pings of each user are clustered first.
	Only those clusters (i.e., not a single trajectory point) with a duration of at least one minute are counted as a ``visit'' \cite{safegraphwhite}.
	By doing so, the daily mobile phone-based visitor flows between CBG and CBG are grouped and summed up.
	A sample record of the daily CBG to CBG visitor flows is included in Table \ref{tab:daily_cbg2cbg}.
	
	For the \textit{weekly CBG to POI visitor flows} metric, different from the \textit{daily CBG to CBG visitor flows metric} which aggregates visitors between origin CBG and destination CBG directly, it provides a mapping of CBGs to POIs.
	In other words, the number of unique visitors who live inside the origin CBG and visit the destination POI in one week are counted.
	A sample record of the weekly CBG to POI visitor flows is also included in Table \ref{tab:daily_cbg2cbg}.
	
	\subsection*{Multiscale Aggregation}
	The two mobile phone-based visitor flows metrics (from CBG to CBG and from CBG to POI) are both processed at the CBG scale. 
	After obtaining these two metrics, all data are further aggregated into three different spatial scales: census tract, county, and state.
	The motivations for providing data products at multiple spatial scales are discussed as follows.
	First, using coarser/finer analysis unit may lead to different outputs, which is known as the scale effect.
	As an important and fundamental concept in geography, the scale effect exists in almost all geographic phenomena \cite{atkinson2000spatial,chen2019quantifying}, including human mobility patterns \cite{liu2012understanding}.
	Providing a multiscale flow dataset allows us to have a more comprehensive view of human mobility and spatial interaction patterns.
	Second, various research projects may acquire data at different spatial scales, depending on the usage.
	For example, for research focusing on a macro view of spatial interactions, state or county scale data might be more suitable as it reflects general regional mobility patterns, while census tract scale data can be used for describing a micro view of human movement patterns such as within cities.
	Also, considering the data size-accuracy trade-off (i.e., the higher the spatial resolution, the higher the accuracy but the larger the data size is), providing a multiscale dataset enables users to download the data that fit their own needs.
	Third, although the daily CBG to CBG visitor flows and weekly CBG to POI visitor flows have been computed at the CBG scale, which is a finer spatial scale, aggregating them to coarser-level can preserve the data privacy better.
	Therefore, the O-D flow dataset is generated at three geographical scales, respectively.
	To do so, we assign census tract, county, and state's geographically unique identifier to each origin CBG and destination CBG that they belong to, and group all records according to the O-D pairs.
	In total, there are 74,001 census tracts, 3,219 counties, and 50 states \& Washington D.C. \& Puerto Rico at each spatial scale for aggregation in the U.S.
	The aggregated mobile phone-based \textit{weekly CBG to POI visitor flows} and the \textit{daily CBG to CBG visitor flows} are termed as \textit{weekly visitor flows} and \textit{daily visitor flows} hereafter, respectively.
	
	\subsection*{Infer Dynamic Population Flows}
	The above-mentioned visitor flows at the three spatial scales are based on mobile phone users detected by SafeGraph, not on the entire population.
	These users account for about 10\% to the entire population in the U.S., and the sampling ratio of unique mobile devices to population varies across CBGs to CBGs \cite{safegraphbias}.
	Existing studies have shown that a good representative sample of the entire population can reflect general human mobility patterns \cite{gonzalez2008understanding}. Also, short-term mobility and long-term migration are affected by multiple factors such as physical travel costs (distance, time, and money), economic and health benefits, social and political frictions  \cite{prieto2018gravity,liu2014uncovering,benach2011migration,fiorio2017using,andris2016integrating,yeghikyan2020learning}. Given that SafeGraph' samples are highly correlated with the true Census populations regarding several socio-economic attributes \cite{safegraphbias}, we aim to infer the short-term population-level dynamic mobility flows during the COVID-19 pandemic as it is important to accurately estimate meta-population infection cases and in other human mobility applications \cite{li2020substantial,lai2020effect,barbosa2018human}.
	To do so, by utilizing the official ACS population data with mobile phone visitor patterns, the dynamic population flows are inferred using the following equation:\\
	\begin{equation}
	pop\_flows(o, d) = visitor\_flows(o, d) \times \frac{pop(o)}{num\_devices(o)}
	\end{equation}
	where $pop\_flows(o, d)$ is the estimated dynamic population flows from geographic unit $o$ to geographic unit $d$, $visitor\_flows(o, d)$ is the computed mobile phone-based visitor flow from $o$ to $d$, $pop(o)$ indicates the population at the geographic unit $o$ extracted from the ACS,  and $num\_devices(o)$ refers to the number of unique mobile devices residing in $o$. In addition, we also compare the estimation results with a gravity model and a radiation model (see more details in the ``Technical Validation'' section).
	
	\section*{Data Records}
	We have produced two data products: weekly flow data and daily flow data, both of which are provided at the census tract, county, and state scales, starting from March 1st, 2020.
	A static copy of the dataset has been uploaded on Figshare \cite{geods_usflow}, while the live version of the dataset will be kept up-to-date with new data stream and can be downloaded from an open data repository on Github: \url{https://github.com/GeoDS/COVID19USFlows}.
	Data provided in this repository are separated into two folders \textit{daily\_flows} and \textit{weekly\_flows} to store daily flow data and weekly flow data, respectively.
	The two folders are organized according to the geographic scale, where \textit{ct2ct} indicates flows between census tract to census tract, \textit{county2county} refers to flows between county to county, and \textit{state2state} contains flow data that originate from one state to others.
	All files are stored in a comma-separated values (CSV) format, which has been widely used for storing, transferring, and sharing data publicly.
	File names are formatted as \{data\_type\}\_\{spatial\_scale\}\_\{date\}.csv, e.g., \textit{weekly\_ct2ct\_03\_02.csv} and \textit{daily\_state2state\_04\_19.csv}.
	Specifically, for weekly flow data, the dates in file name refers to the date of the Monday in that week but summarize all mobility flows in that week from Monday to Sunday.
	Since the file size of flow data at the census tract scale exceeds the GitHub disk limit, each flow data file is split into 20 subfiles (that can be merged after downloading).
	The daily and weekly aggregated directional flows are coded by pairwise origin to destination units using geo-identifiers (GEOIDs). 
	The GEOIDs are numeric codes that uniquely identify different administrative levels (e.g., census tract, county, and state) in the U.S. Census Bureau data portal \cite{geoids}. 
	For the GEOIDs at each scale, census tract is using an 11-digit number, county is a 5-digit number, and state is a 2-digit number.
	The coordinates of origins and destinations can be used for creating spatial interaction flow maps. 
	Reference shapefiles at each scale are available from TIGER/Line Shapefiles (\url{https://www.census.gov/geographies/mapping-files/time-series/geo/tiger-line-file.html}).
	External demographic and socioeconomic statistical information at different spatial scales can be accessed directly from the U.S. Census Bureau and joined to each origin and destination using GEOIDs (\url{https://www.census.gov/data.html}).  
	A description of all attributes in the dataset is shown below.

	\subsubsection*{Weekly Flow Data}
	geoid\_o - Unique identifier of the origin geographic unit (census tract, county, and state). Type: string. \\
	geoid\_d - Unique identifier of the destination geographic unit (census tract, county, and state). Type: string.\\
	lat\_o - Latitude of the geometric centroid of the origin unit. Type: float.\\
	lng\_o - Longitude of the geometric centroid of the origin unit. Type: float.\\
	lat\_d - Latitude of the geometric centroid of the destination unit. Type: float.\\
	lng\_d - Longitude of the geometric centroid of the destination unit. Type: float.\\
	date\_range - Date range of the records. Type: string.\\
	visitor\_flows - Estimated number of visitors detected by SafeGraph between the two geographic units (from geoid\_o to geoid\_d), computed and aggregated from weeekly CBG to POI flows. Type: float.\\
	pop\_flows - Estimated entire population flows between the two geographic units (from geoid\_o to geoid\_d), inferred from visitor\_flows as described in the subsection `Infer Dynamic Population Flows'. Type: float.\\

	\subsubsection*{Daily Flow Data}
	geoid\_o -  Unique identifier of the origin geographic unit (census tract, county, and state). Type: string. \\
	geoid\_d - Unique identifier of the destination geographic unit (census tract, county, and state). Type: string.  \\
	lat\_o - Latitude of the geometric centroid of the origin unit. Type: float.\\
	lng\_o - Longitude of the geometric centroid of the origin unit. Type: float.\\
	lat\_d - Latitude of the geometric centroid of the destination unit. Type: float.\\
	lng\_d - Longitude of the geometric centroid of the destination unit. Type: float.\\
	date - Date of the records. Type: string.\\
	visitor\_flows - Estimated number of visitors between the two geographic units (from geoid\_o to geoid\_d), computed and aggregated from daily CBG to CBG flows. Type: float.\\
	pop\_flows - Estimated entire population flows between the two geographic units (from geoid\_o to geoid\_d), inferred from visitor\_flows as described in the subsection 'Infer Dynamic Population Flows'. Type: float.\\

	\section*{Technical Validation}
	To check the data distribution and ensure the reliability of the produced mobility O-D flow dataset, three complementary methods are employed for data validation.
	We first compare the probability distributions of the computed visitor flows and the estimated entire population flows to check if the data distributions stay consistent during the production process.
	Then, we compare the estimated population flows using our inference approach with results estimated from a gravity model and a radiation model. Lastly, the released mobility flow dataset is compared with two other openly available data sources: ACS commuting flows and the Descartes Labs mobility changes.
	The hypothesis is that mobility patterns at the same geographic scale should be consistent across multiple data sources.
	
	\subsection*{Checking Distributions of Visitor Flows and Population Flows}
\begin{figure}[H]
		\centering
		\includegraphics[width=\textwidth]{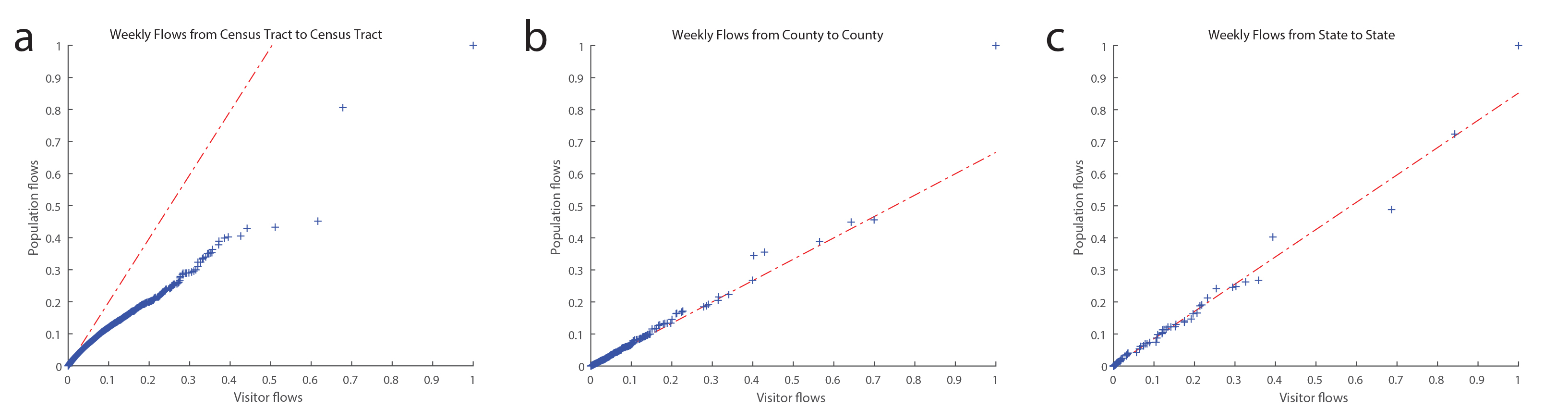}
		\caption {Quantile-Quantile plots of visitor flows and population flows based on weekly flow data in the week of March 2nd to March 8th, 2020. (a) at the census tract scale; (b) at the county scale; and (c) at the state scale. The red lines are fitting lines between two distributions.} 
		\label{fig:qqplot}
	\end{figure}

	Following the methods described above, the two types (daily and weekly) of mobile phone-based visitor flows are directly aggregated at different geographic scales, while the demographic data are involved in estimating the entire population flows.
	Visitor flows and population flows are supposed to have similar distributions, and thus to make sure the inferring process keeps the distributions of the mobility flows unchanged, we made the Q–Q (quantile-quantile) plots to compare their distributions.
	Visitor flows and population flows are first normalized, as they have different value ranges.
	If the distributions of the two metrics being compared are linearly related, the scatter points should locate following a line $y = kx$, where $x,y$ are percentiles of the two datasets and $k$ is the coefficient.
	Figure \ref{fig:qqplot} shows the Q-Q plots of the visitor flows and population flows at three spatial scales (a: census tract, b: county, and c: state) based on the weekly flow data in the week of 03/02/2020 to 03/08/2020.
	Though we only plot the two distributions using one week data as an example, the associations between visitor flows and population flows are similar for other dates.
	As is shown in Figure \ref{fig:qqplot}, scatter points are distributed along a straight line at both county scale and state scale. Even though the line is not $y = x$, the inferred entire population flows are linearly related to the mobile phone-based visitor flows ($R^2$ is 0.958 at county scale and 0.953 at state scale) and keep consistent distributions (Figure \ref{fig:qqplot}b and \ref{fig:qqplot}c).
	The distribution was slightly different at census tract scale. 
	Though most scatter points are distributed following the fitting line ($R^2$ is 0.865 at census tract scale), those points with relatively high visitor flows and populations flows are located below the fitting line (Figure \ref{fig:qqplot}a).
	The reason is that most paired census tracts have only a few visits and scatter points aggregate near the coordinate origin (and these points dominate the fitting line slope parameter).
	Therefore, the distributions of visitor flows and population flows keep consistency especially for those paired regions with a small number of flows but have larger uncertainty for regions with large number of flows.
	In sum, it can be concluded that the computed mobile phone-based visitor flows and the inferred entire population flows generally have linear associations across different spatial scales. Since the overall flow distributions are similar and generally consistent across geographic scales, the population flows inference process is reliable.

	\subsection*{Comparison with Gravity Model and Radiation Model Estimations}
	
	The gravity-style and radiation-style models that have been widely used in modeling spatial interaction patterns and human mobility O-D flows \cite{simini2012universal,kang2015generalized,prieto2018gravity}. Thus, we also employ a classic gravity model and a radiation model to estimate the visitor/population flows between counties and then examine the correlation between the model outputs and our dataset estimates.
	The comparison experiments are performed at the county scale using the weekly flow data considering that most of COVID-19 open data are available at the county scale in the U.S. \cite{dong2020interactive,killeen2020county}.
	
	The gravity model assumes that the magnitude of the flow between two places is related to their nodal attraction and in reverse proportion to the distance between them as shown by the following equation \cite{o1995new}. 
	
	\begin{equation}
	F_{ij} = \frac{k\times P_{i}\times P_{j}}{d_{ij}^\beta}
	\end{equation}
	
	\noindent where $F_{ij}$ represents the flow between county $i$ and $j$; $P_i$ and $P_j$ are the total number of mobile devices (or the population) in the county $i, j$ respectively; $d_{ij}$ is the distance between the two counties. $k$ and $\beta$ are parameters that are used to adjust the model to fit the observations. For each week's data, we use the Particle Swarm Optimization technique \cite{eberhart1995new,liang2020calibrating} to find the best parameter combination of $k$ and $\beta$ that minimize the RMSE (Root Mean Square Error) between the estimated flow and the directly computed mobile phone-based visitor flow for that week. To make sure the estimated flow and the visitor flows are at the same scale, the range of $k$ is set to be between 0 to 0.001. The range of $\beta$ is between 0 and 2. The gravity model parameter fitting results for each week between 03/02/2020 and 05/31/2020 are shown in Table \ref{tab:gravityraditionmodels}. The value of $\beta$ ranges from 0.85 to 1.00. After the statewide lockdowns in the U.S. \cite{gao2020mobile}, the distance decay coefficient $\beta$ increased slightly, which showed that there were fewer longer-distance trips compared to that before the pandemic.
	Then we applied the parameters in the census population to estimate the population flow of each week. 
	The Pearson's correlation coefficient between the gravity model output and the population flow estimates (pop\_flows) is about 0.56 to 0.64.
	
	One problem with the gravity model is that it can only capture undirected spatial interactions as the origin and the destination are not specified.
	So we also use the radiation model that considers the direction of the flow for estimation. The original radiation model is used to estimate the commuting flows based on the U.S. census population distribution \cite{simini2012universal}. Here we applied this model to estimate the visitor flows based on the number of mobile phone users that travel across counties, the flow $T_{ij}$ from county $i$ to county $j$ can be computed as follows:
	
	\begin{equation}
	T_{ij} = T_i\frac{m_i \times n_j}{(m_i + s_{ij})\times(m_i + n_j + s_{ij})}
	\end{equation}
	\noindent where $m_i$ is number of mobile phone devices in the source county $i$; $n_j$ is number of mobile phone devices in the destination county $j$; $s_{ij}$ is the total number of mobile phone devices in the counties that are within the circle of radius $r_{ij}$ (distance between $i, j$) centered at county $i$ (exclude $i, j$). $T_i$ represents all the visitor flows that come from county $i$, and the model assumes that this is proportional to the total number of movement population of the source county \cite{simini2012universal}. Therefore, we have
	\begin{equation}
	T_i = m_i(N_c/N)
	\end{equation}
	
	\noindent where $N_c$ is considered as the total number of devices that move (having at least one trip) during the study period, $N$ is the total number of devices. In our daily data, we have the attribute that indicates the number of devices that are completely at home during that day. Therefore, we used the subtraction of the total number of devices and the number of `completely at home' devices as the $N_c$. The weekly $N_c$ is the average of the 7-day's $N_c$ values. For each week, we estimated the flow $T_{ij}$ for all counties and computed the Pearson's correlation between the estimated flow and the `pop\_flows' in our dataset (in Table \ref{tab:gravityraditionmodels}). The correlation between the radiation model estimates and the `pop\_flows' is about 0.75, which is better than that of the gravity model estimates.
	
	As there is no ground-truth, such a comparison with the gravity model and the radiation model provides a cross-referencing aspect and can inform the potential data users about their differences. 

	\subsection*{Comparison with Other Data Sources}
	To illustrate that the data quality is high, we then compare two openly available data sources to the produced dataset from two different aspects, namely O-D flow patterns and temporal patterns.
	Correlation analysis is conducted to check if mobility patterns of the compared two datasets differ.
	Though there is no ground-truth data which characterizes the real dynamic population flows between two geographic regions as of yet, the comparison results are still useful for evaluating the credibility of the produced mobility flow dataset.
	
	In terms of the O-D flow patterns between two regions, we take the \textit{ACS commuting flows} as baseline to see if the mobility patterns before the pandemic detected in our data products have a high correlation with the ACS commuting flows patterns \cite{commuting_flow}.
	ACS commuting flows are generated by asking participants about their residence locations and primary workplace locations, and such an O-D flow dataset informs the understanding of interconnectedness between communities.
	ACS commuting flows provide data at county (and county subdivision) scales \cite{commuting_flow} and thereby our comparison experiments are conducted only at the county scale.
	The ACS commuting flows reflect the general patterns of commuting flows in the U.S., i.e., normal flows not affected by the COVID-19 pandemic.
	Due to the fact that a national emergency concerning the COVID-19 pandemic was declared in the U.S. on March 13, 2020 and followed by statewide stay-at-home orders \cite{gao2020mobile}, three time slices are picked up for comparison as they are supposed to reflect the mobility patterns before, during, and after the stay-at-home orders, respectively.
	For daily flow data, we chose March 2nd, April 6th, May 11th, and May 25th as examples for comparison, while for weekly flow data we chose the weeks of March 2nd-8th (before the orders), April 6th-12th (during the orders), May 11th-17th (several states reopened after lifting the orders), and May 25th-31st  (business reopened in most states after the Memorial day \cite{nyt_reopen}) for comparison.
	
	Comparison results between our produced mobility flow dataset and the ACS commuting flows data are illustrated in Table \ref{tab:corr_acs}.
	Regarding the number of records, the ACS commuting flows data contains 137,806 rows, while the generated mobility flow dataset at the county scale contains more origin to destination pairs than the ACS commuting flow data.
	Our mobility flow dataset has a higher quantity of spatial interactions between county to county as more types of flows (including not just the home-to-job commuting flow) are captured.
	The total number of records before the stay-at-home orders is greater than the number of records during and after stay-at-home orders, which is an intuitive consequence of people reducing their movements during the pandemic.
	After joining the two datasets and removing records with no O-D matches, as is shown in Table \ref{tab:corr_acs}, both weekly flow data and daily flow data had a high agreement (with greater than 0.93 Pearson's correlation) with the ACS commuting flow data.
	In particular, the inferred entire population flows have higher correlation coefficients with ACS commuting flows than the directly computed mobile phone-based visitor flows.
	With respect to the temporal changes, the flow patterns before stay-at-home orders have higher correlation coefficients, which matches expectations as people reduced their mobility when the stay-at-home orders in place \cite{gao2020mapping}, leading to a decrease in the correlation coefficient values.
	The comparison results illustrate that our proposed dataset may be complementary to the official ACS commuting flow data.
	In addition, different from the ACS commuting flow data which only provides every 5-year updates at county scale and at minor civil division scale in major metropolitan areas, our dataset enables researchers to explore human mobility patterns at multiple spatial scales and in higher temporal resolutions (i.e., daily and weekly time windows).
	
	Additionally, the temporal patterns of the introduced mobility flow dataset and the Descartes Lab's COVID-19 mobility changes dataset is compared.
	Descartes Lab's mobility changes dataset provides a mobility index which shows the median of the max-distance mobility of all users detected in one specific region \cite{descarteslabs2020mobility}.
	Such an index has been widely used for monitoring the daily mobility changes in the U.S. \cite{gao2020mapping}. These two datasets are assumed to have similar temporal trends.
	To do so, the total number of mobile phone-based visitor flows and entire population flows in the daily mobility flow dataset and their mobility changes are matched in five U.S. metropolitan areas: New York, Los Angeles, Chicago, Houston, as they are the top four metropolitan areas with largest population in the U.S., and Seattle, where the first confirmed COVID-19 case was reported \cite{holshue2020first}.
	Correlation analysis is performed to check if these two datasets capture similar mobility changes during the epidemic since March 1st, 2020.
	
	As is shown in the Table \ref{tab:corr_des}, the produced mobility flow datasets have high correlation coefficients (at least 0.92) with the Descartes Lab's mobility changes dataset in all five metropolitan areas.
	It shows that the generated datasets capture similar mobility temporal patterns with other open data sources.
	As mentioned above, there is no ground-truth in measuring such high resolution dynamic human mobility patterns for the entire population, and as such it is therefore impossible to know which one can characterize human mobility more accurately.
	However, by cross-referencing other sources, high correlation coefficients indeed show the reliability of our generated mobility flow dataset.

	\section*{Usage Notes}
	Different datasets have their own pros and cons. 
	In this data descriptor, we introduce two types of datasets in characterizing human dynamic O-D flows: daily flows and weekly flows, each at three geographic scales (i.e., census tract, county, and state).
	A detailed description of the statistics and distributions of the datasets may benefit different applications.
	Here, we discuss the characteristics and limitations of our mobility flow dataset to guide potential usages.
	
\begin{figure}[H]
		\centering
		\includegraphics[width=\textwidth]{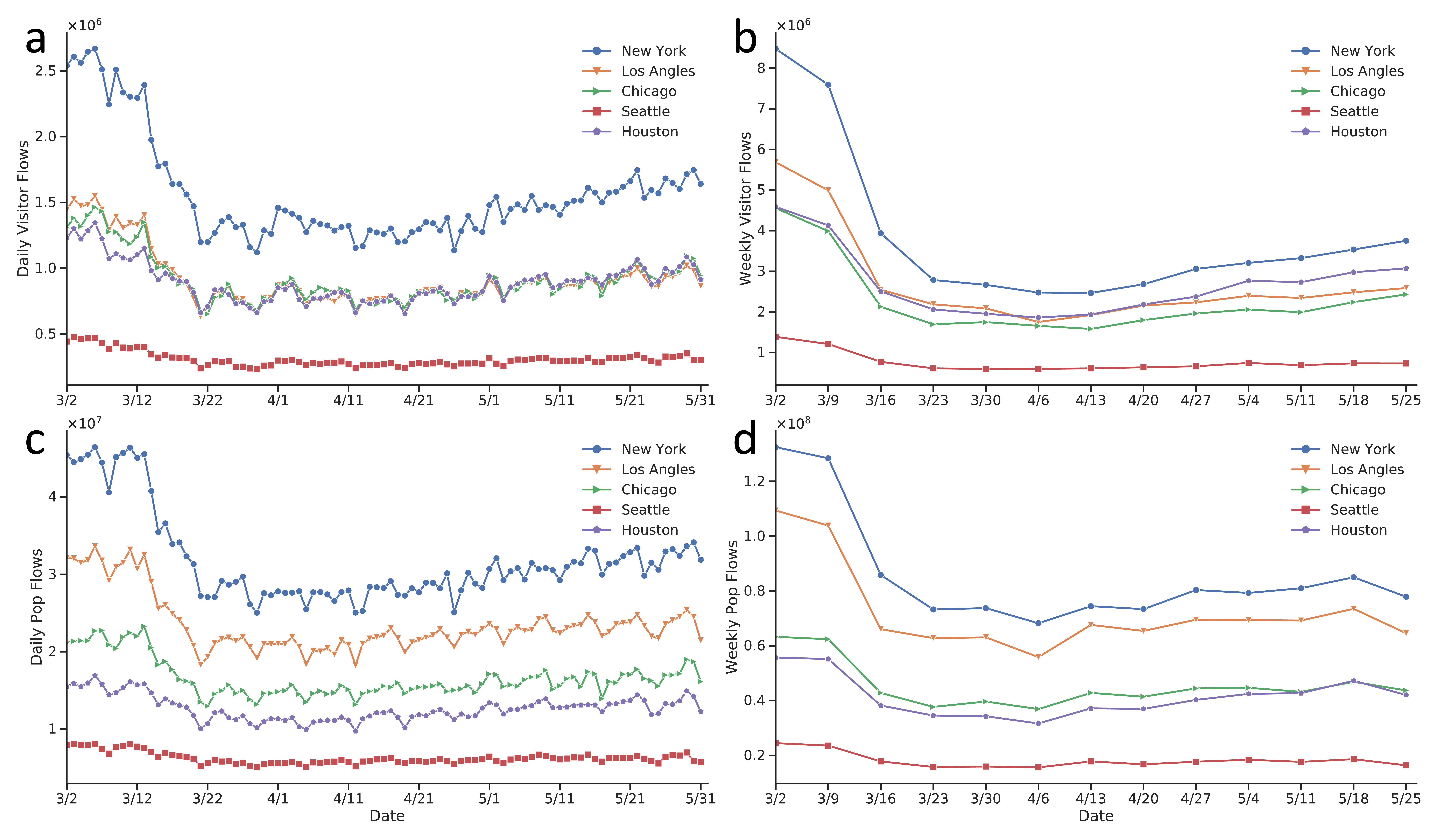}
		\caption{Temporal patterns of mobility flows in five metropolitan areas: New York, Los Angeles, Chicago, Seattle, and Houston. (a) daily visitor flows; (b) daily population flows; (c) weekly visitor flows; (d) weekly population flows. Date range: from March 2nd to May 31st, 2020.}
		\label{fig:temporal_patterns}
	\end{figure}

	Figure \ref{fig:temporal_patterns} shows the temporal changes of the total number of mobility flows in five metropolitan areas: New York, Los Angeles, Chicago, Houston, and Seattle.
	Accordingly, the daily flow data provides a more detailed temporal pattern description of human mobility changes compared to the weekly flow data. 
	As illustrated in Figure \ref{fig:temporal_patterns}, the temporally-changing curves have more fluctuations over time as human mobility patterns might be influenced by various factors and individual events (e.g., statewide lockdowns, presidential primary election day, Memorial day, or street protests), while weekly flow data reflects more general mobility patterns;
	the temporal curves for the weekly flow data are more smooth and stable in comparison to the daily flow data.
	For example, people may not visit the supermarket everyday and therefore the visitor volumes may vary by day of a week, while the sum of volumes in one week is more stable.
	
\begin{figure}[H]
		\centering
		\includegraphics[width=\textwidth]{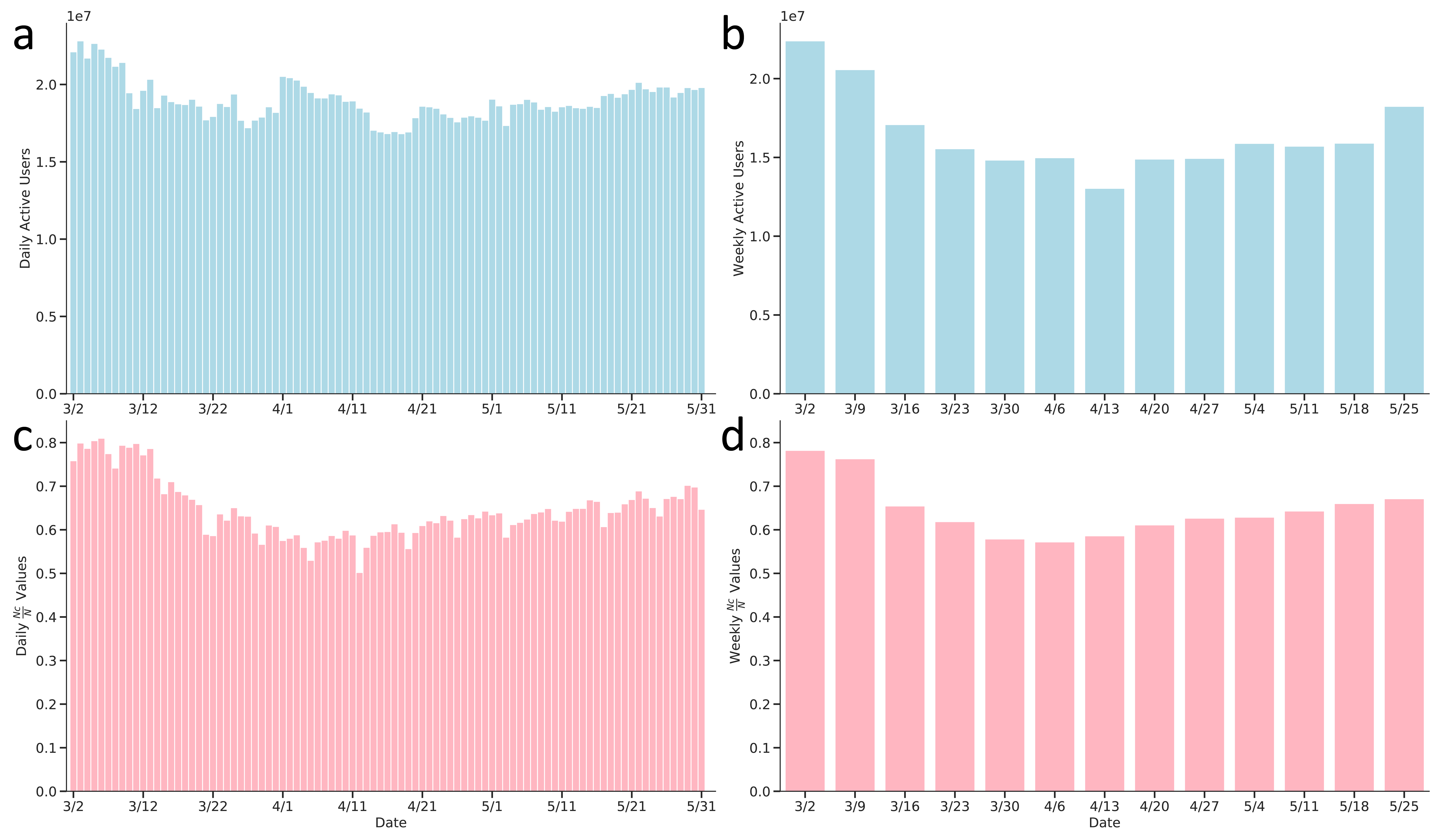}
		\caption{The number of active mobile phone users and the Nc/N value over time, where Nc represents the number of users that have at least one trip, and N represents the total number of mobile phone users observed in each period. (a) daily active users; (b) weekly active users; (c) daily Nc/N values; (d) weekly Nc/N values. Date range: from March 2nd to May 31st, 2020.}
		\label{fig:temporal_active_users}
	\end{figure}

	Figure \ref{fig:temporal_active_users} shows the temporal variations of daily and weekly active user sample size and the ratio of users having at least one trip ($N_c/N$) respectively. We find that the number of active users and the $N_c/N$ values change over time and have a similar changing trend to the total number of mobility flows. One would expect fewer active users to be detected after the stay-at-home mandates as more people stayed at home more frequently.
	The weekly number of active devices detected in the dataset reached over 22 million at the beginning and decreased to a minimum of 13 million active users in the week of April 13-19, 2020.
	Similarly, as the outbreak progressed, the daily number of active users identified in the dataset decreased to a minimum of 16 million residing users on April 18.
	After that, more active users were detected due to reduced compliance with stay-at-home mandates and the eventual lifting of stay-at-home orders.
	
\begin{figure}[H]
		\centering
		\includegraphics[width=\textwidth]{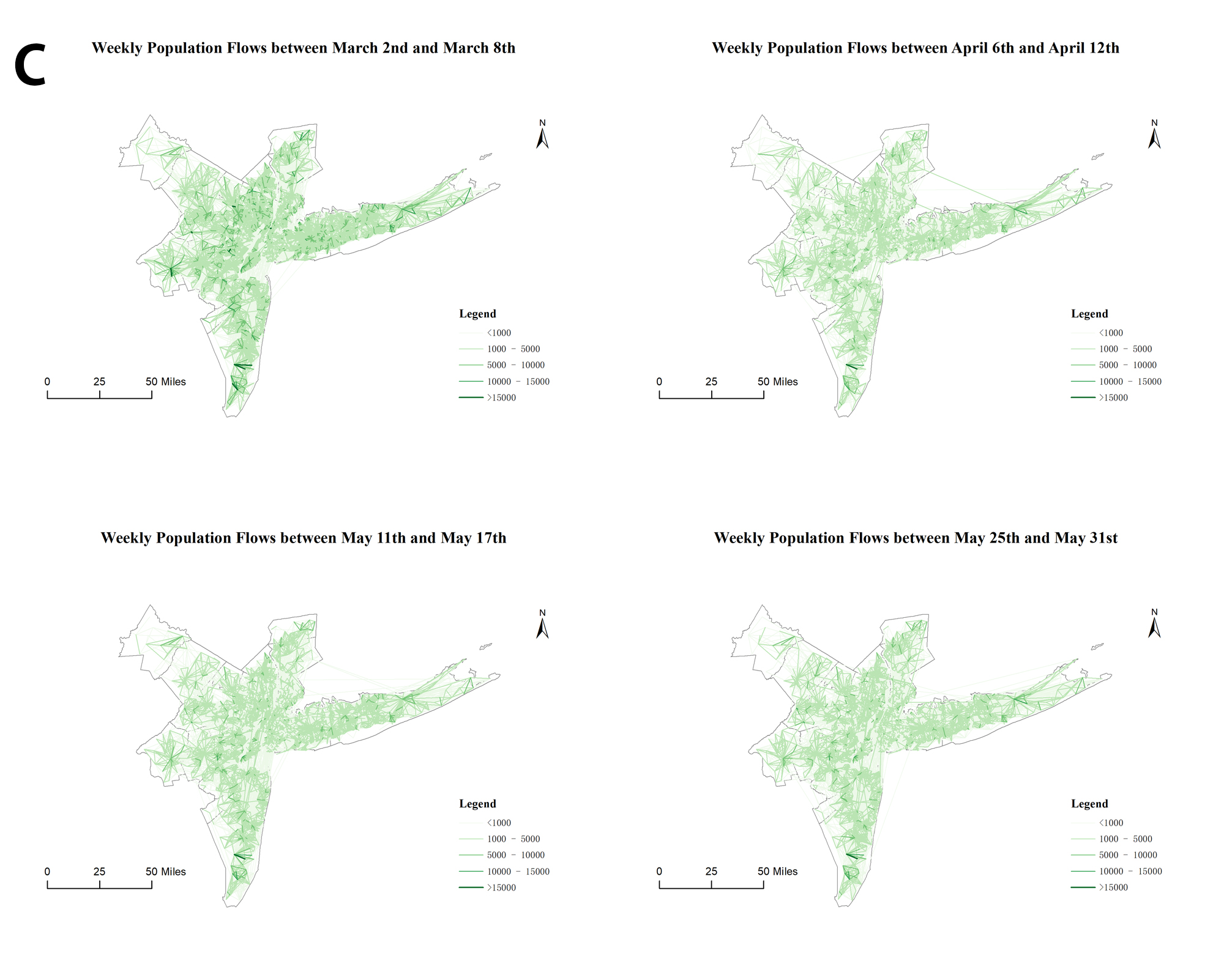}
		\caption{Spatial patterns of mobility flows before (March 2nd to March 8th), during (April 6th to April 12th), and after (May 11th to May 17th, May 25th to May 31st) the stay-at-home orders at three geographic scales using weekly flow data. A: From state to state across the Contiguous U.S.; B. From county to county across the Contiguous U.S.; C. From census tract to census tract in the New York metropolitan area. Note that the flow dataset includes all 50 states, Washington D.C. and Puerto Rico; the flows in Hawaii, Alaska, and Puerto Rico are not shown in the map.}
		\label{fig:flow_maps}
	\end{figure}

	Figure \ref{fig:flow_maps} shows the spatial patterns of O-D flow changes during the COVID-19 pandemic based on the weekly flow mobility data.
	Figure \ref{fig:flow_maps}a and \ref{fig:flow_maps}b show the spatial interaction patterns of population flows across the Contiguous U.S. at the state and county scales, respectively. 
	Figure \ref{fig:flow_maps}c shows the population flow patterns at the census tract scale in the New York metropolitan area (New York was selected as it had the most confirmed COVID-19 cases during the study period).
	We take four weekly flow data to represent the spatial patterns of mobility flows before (March 2nd to March 8th), during (April 6th to April 12th), and after (May 11th to May 17th and May 25th to May 31st) the stay-at-home orders.
	At all three spatial scales, the movement flows decrease significantly from March to April due to the stay-at-home orders, with certain increases in May with the start of state partial reopenings. The decrease and subsequent increase in human movement flows clearly show how the outbreak of the COVID-19 pandemic and social distancing-related policies affect human mobility changes.
	In particular, according to Figure \ref{fig:flow_maps}b, during the stay-at-home order period long-range spatial interactions decrease to a small quantity, while most human movements appear as short-range movements to the adjacent counties.
	When cities begin to reopen in May, long-term spatial interactions gradually rebound at both state scale and county scale.

	While we made great efforts to guarantee the reliability of our produced mobility flow datasets and reduce the data uncertainty, a few limitations should be acknowledged.
	
	First, since we acquire anonymous mobile phone data from SafeGraph, the privacy policies applied should be considered by the end-user, as they may influence data uncertainty.
	The weekly flow data is derived from CBG to POI visitor flow metric.
	Please note that when a CBG has no visitors or has only one visitor who originate from that CBG to one POI, the visit between CBG to POI will not be recorded.
	If the CBG has two to four visitors who originate from that CBG to another POI, the visitor count will be recorded but shown as four to enhance differential privacy.
	Only when the CBG has more than five visitors whose home places inside the CBG to one POI will the real visitor count be displayed correctly.
	Consequently, the least number of visitors from a CBG to POI as well as the weekly flow data computed in one week is four.
	In addition, only when a visitor stays at a place (CBG or POI) for more than one minute will  a ``visit'' be counted. Such a relatively short time period may also affect the foot-traffic counting and cause data uncertainty.

	Second, in terms of foot-traffic counting criteria, it should be noted that, for the daily flow data, it records the unique visitors to one CBG in one day, while for the weekly flow data it records the unique visitors to one POI in one week.
	Due to the different counting methods, the sum of daily flow data in one week (i.e., seven days) are not equal to the weekly flow data. 
	For the daily flow data, which is computed directly based on the daily CBG to CBG visitor flows, it may underestimate the mobility flows to POIs within one CBG.
	That is, a visitor may visit more than one POI inside that CBG (e.g., different buildings in a campus) during the same day, it would still only be recorded once from the origin CBG to the destination CBG.
	For the weekly flow data, which is based on the weekly CBG to POI visitor flows, mobility flows to POIs in one week may be underestimated as a visitor may visit one POI multiple times (e.g., work places) in one week, which would only be counted once as we compute the unique visitor devices rather than raw visits.
	
	In addition, visitor duplication may also exist when aggregating the flows from lower level spatial scales to upper level spatial scales (e.g., from CBG level to census tract scale, county scale, or state scale), and it may lead to the inflation of the inferred dynamic population flows.
	For example, a user may originate from his/her home CBG and visit two POIs (e.g., one coffee shop and one grocery store) which are located inside different CBGs but at the same census tract.
	Such a visitor should be counted as one unique visitor from the home census tract to the POI census tract. 
	However, when aggregating the visits from CBG to CBG mobility flow to census tract to census tract mobility flow, because the individual travel behaviour cannot be traced, the mobility flows between these two census tract are counted as two and thereby inflate the real mobility flows.
	
	Last but not least, data bias is a common issue for large-scale mobile phone data that may influence the representativeness of our produced dataset.
	While dynamic mobility flows are inferred from mobile phone applications by users, not everyone in the population has a mobile phone, and not everyone uses smartphone applications, especially elderly people and children. 
	Given these differences in mobile phone usage, age groups and demographic composition might influence the estimated entire population mobility flows. With these caveats in mind, this data still provides important up-to-date mobility flow information at scale.
	
	In conclusion, such a timely-produced dynamic O-D human flow dataset at different geographic scales can help deepen our understanding of human dynamics under this public health crisis, inform public health policy making, and support many other social sensing and transportation applications. 
	
	
	\section*{Code availability}
	Data processing and data analysis were performed on a Linux server using the Python version 3.7.
	All codes used for analysis are available in the public GitHub repository that hosts the data: \url{https://github.com/GeoDS/COVID19USFlows}. 
	
	\section*{Acknowledgements}
	We would like to thank the SafeGraph Inc. for providing the anonymous and aggregated place visit data. S.G. acknowledges the funding support provided by the National Science Foundation (Award No. BCS-2027375). Any opinions, findings, and conclusions or recommendations expressed in this material are those of the author(s) and do not necessarily reflect the views of the National Science Foundation. Support for this research was partly provided by the University of Wisconsin - Madison Office of the Vice Chancellor for Research and Graduate Education with funding from the Wisconsin Alumni Research Foundation.
	
	\section*{Author contributions}
	Research design and conceptualization: S.G., Y.K.; Data collection and processing: Y.K., S.G.; Result analysis: Y.K., S.G., Y.L.; Visualization: M.L., J.R., Y.K., S.G.; Writing: all authors. 
	
	
	\section*{Competing interests}
	
	The authors declare no competing interests.
	
	\section*{Figures}
	
	\begin{figure}[H]
		\includegraphics[width=\textwidth]{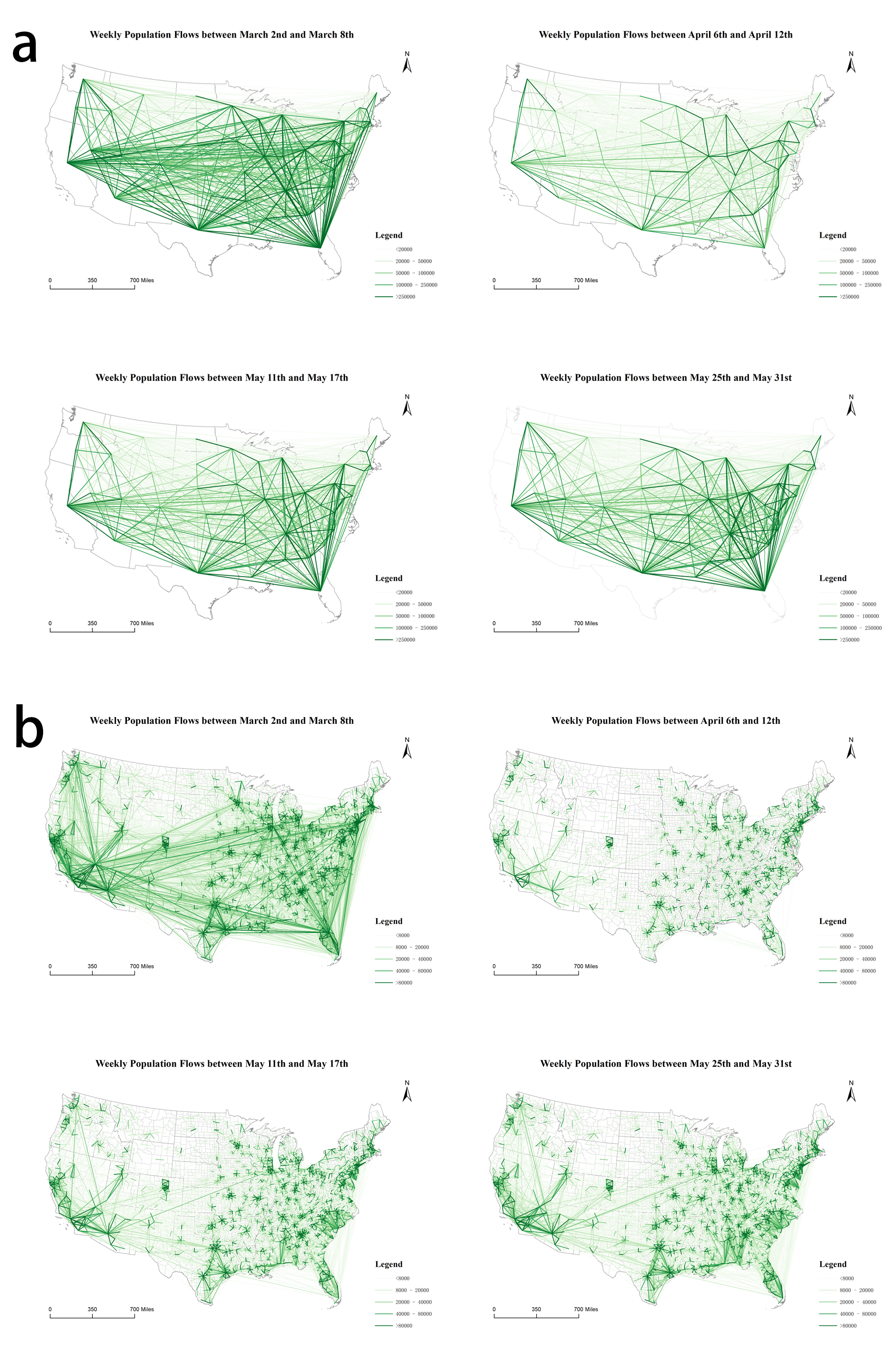}
	\end{figure}

	

	\section*{Tables}
	
	\begin{table}[H]
		\centering
		\caption{A sample record of: (A) daily CBG to CBG visitors; (B) weekly CBG to POI visitors.}
		\begin{tabular}{@{}llll@{}}
			\toprule
			Type                           & Origin       & Destination  & Visitors \\ \hline
			(A) daily CBG to CBG visitors  & 0123456789xx & 0123456798xx & 50       \\
			(B) weekly CBG to POI visitors & 0123456789xx & sg:012345xx  & 10       \\ \bottomrule
		\end{tabular}
		\label{tab:daily_cbg2cbg}
	\end{table}
	

	\begin{table}[H]
		\centering
		\caption{The gravity model and the radiation model parameter settings and the correlation between the model output and the population flow estimates.}
		\label{tab:gravityraditionmodels}
		\begin{tabular}{@{}llllll@{}}
			\toprule
			\multirow{2}{*}{Date} & \multicolumn{3}{l}{Gravity Model}   & \multicolumn{2}{l}{Radiation Model} \\ \cmidrule(l){2-6} 
			& k         & $\beta$      & correlation & Nc/N          & correlation         \\\hline 
			03-02                 & 0.000049300 & 0.8636853 &0.6484     & 0.782         & 0.755               \\
			03-09                 & 0.000062300 & 0.9010593 &0.6301     & 0.763         & 0.751               \\
			03-16                 & 0.000065800 & 0.9419980 &0.6108     & 0.654         & 0.756               \\
			03-23                 & 0.000070300 & 0.9505486 &0.5852     & 0.619         & 0.753               \\
			03-30                 & 0.000078200 & 1.0023736 &0.5698     & 0.579         & 0.748               \\
			04-06                 & 0.000072100 & 0.9754115 &0.5656     & 0.572         & 0.749               \\
			04-13                 & 0.000077600 & 0.9297287 &0.5620     & 0.586         & 0.747               \\
			04-20                 & 0.000090900 & 1.0044105 &0.5602     & 0.611         & 0.753               \\
			04-27                 & 0.000076800 & 0.9284104 &0.5682     & 0.627         & 0.756               \\
			05-04                 & 0.000080000 & 0.9269356 &0.5690     & 0.629         & 0.757               \\
			05-11                 & 0.000061200 & 0.8748065 &0.5721     & 0.643         & 0.756               \\
			05-18                 & 0.000060800 & 0.8532109 &0.5718     & 0.660         & 0.758               \\
			05-25                 & 0.000061600 & 0.9175823 &0.5645     & 0.671         & 0.756               \\ \bottomrule
		\end{tabular}
	\end{table}

	\begin{table}[H]
		\centering
		\caption{Pearson's correlation coefficients between our mobility flow dataset and the ACS commuting flow dataset at county scale.}
		\resizebox{\textwidth}{!}{
			\begin{tabular}{@{}llllllll@{}}
				\toprule
				\multicolumn{4}{l}{Weekly   Flow Data}                                                   & \multicolumn{4}{l}{Daily Flow Data}                                                      \\ \midrule
				Date      & Type             & Matched Records         & Pearson Correlation Coefficient & Date      & Type             & Matched Records         & Pearson Correlation Coefficient \\
				3/2/2020  & Visitor Flows    & \multirow{2}{*}{102750} & 0.961                           & 3/2/2020  & Visitor Flows    & \multirow{2}{*}{102206} & 0.953                           \\
				3/2/2020  & Population Flows &                         & 0.984                           & 3/2/2020  & Population Flows &                         & 0.985                           \\
				4/6/2020  & Visitor Flows    & \multirow{2}{*}{78577}  & 0.932                           & 4/6/2020  & Visitor Flows    & \multirow{2}{*}{92297}  & 0.935                           \\
				4/6/2020  & Population Flows &                         & 0.981                           & 4/6/2020  & Population Flows &                         & 0.98                            \\
				5/11/2020 & Visitor Flows    & \multirow{2}{*}{91069}  & 0.917                           & 5/11/2020 & Visitor Flows    & \multirow{2}{*}{94729}  & 0.934                           \\
				5/11/2020 & Population Flows &                         & 0.977                           & 5/11/2020 & Population Flows &                         & 0.981                           \\ 
				5/25/2020 & Visitor Flows    & \multirow{2}{*}{95350}  & 0.915                           & 5/25/2020 & Visitor Flows    & \multirow{2}{*}{99037}  & 0.934                           \\
				5/25/2020 & Population Flows &                         & 0.974                           & 5/25/2020 & Population Flows &                         & 0.980                           \\ \cmidrule(r){1-3} \cmidrule(lr){4-7} \cmidrule(l){8-8} 
		\end{tabular}}
		\label{tab:corr_acs}
	\end{table}

	\begin{table}[H]
		\centering
		\caption{Pearson correlation coefficients between the temporal patterns of our daily flow dataset and that of the Descartes Labs dataset.}
		\begin{tabular}{@{}ll@{}}
			\toprule
			Metropolitan Area & Pearson Correlation Coefficient \\ \midrule
			New York          & 0.977                   \\
			Los Angeles       & 0.956                   \\
			Chicago           & 0.928                   \\
			Houston           & 0.92                    \\
			Seattle           & 0.951                   \\ \bottomrule
		\end{tabular}
		\label{tab:corr_des}
	\end{table}

	
	

	\bibliographystyle{acm}
	\bibliography{references}
	
	
\end{document}